\documentclass[usenatbib]{mn2e}
\usepackage{natbib}
\usepackage{mathtools}
\input psfig.sty
\usepackage{epsfig}
\usepackage{amsmath,amssymb}
\usepackage{psfig}
\usepackage{graphicx}

\newcommand{\kpc} {{\,\rm kpc}}

\newcommand{\kms}{{\,\rm {km\,s^{-1}} }}

\renewcommand{\vec}[1]{{\mathbf #1}}

\def \apjs  {ApJS}
\def \apjl  {ApJL}

\def \etal {et~al.~}\def \chisq  {\ifmmode  \chi^2   \else  $\chi^2$  \fi}  
\def \spose#1{\hbox  to 0pt{#1\hss}}  
\def \lta{\mathrel{\spose{\lower 3pt\hbox{$\sim$}}\raise  2.0pt\hbox{$<$}}}
\def \gta{\mathrel{\spose{\lower  3pt\hbox{$\sim$}}\raise 2.0pt\hbox{$>$}}}
\def \kms {\ifmmode  \,\rm km\,s^{-1} \else $\,\rm km\,s^{-1}  $ \fi }
\def \kpc {\ifmmode  {\rm kpc}  \else ${\rm  kpc}$ \fi  }  
\def \Msun {\ifmmode M_{\odot} \else $M_{\odot}$ \fi} 
\def \hMsun {\ifmmode h^{-1}\,\rm M_{\odot} \else $h^{-1}\,\rm M_{\odot}$ \fi}
\def \LCDM {\ifmmode \Lambda{\rm CDM} \else $\Lambda{\rm CDM}$ \fi}
\def \sig8 {\ifmmode \sigma_8 \else $\sigma_8$ \fi} 
\def \OmegaM {\ifmmode \Omega_{\rm M} \else $\Omega_{\rm M}$ \fi} 
\def \OmegaL {\ifmmode \Omega_{\rm \Lambda} \else $\Omega_{\rm \Lambda}$\fi} 
\def \Deltavir {\ifmmode \Delta_{\rm vir} \else $\Delta_{\rm vir}$ \fi}
\def \rhocrit {\ifmmode \rho_{\rm crit} \else $\rho_{\rm crit}$ \fi}
\def \rhou {\ifmmode \rho_{\rm u} \else $\rho_{\rm u}$ \fi}
\def \zc {\ifmmode z_{\rm c} \else $z_{\rm c}$ \fi}
\def \rhos {\ifmmode \rho_{\rm s} \else $\rho_{\rm s}$ \fi} 
\def \rs {\ifmmode r_{\rm s} \else $r_{\rm s}$ \fi} 
\def \cvir {\ifmmode c_{\rm vir} \else $c_{\rm vir}$ \fi} 
\def \Rvir {\ifmmode r_{\rm vir} \else $R_{\rm vir}$ \fi}
\def \Vvir {\ifmmode V_{\rm  vir} \else  $V_{\rm vir}$  \fi} 
\def \Mvir {\ifmmode M_{\rm  vir} \else $M_{\rm  vir}$ \fi}  
\def \Nvir {\ifmmode N_{\rm  vir} \else $N_{\rm  vir}$ \fi}  
\def \Jvir {\ifmmode J_{\rm vir} \else $J_{\rm vir}$ \fi} 
\def \Evir {\ifmmode E_{\rm vir} \else $E_{\rm vir}$ \fi} 
\def \lam {\ifmmode \lambda  \else $\lambda$ \fi} 
\def \lamp {\ifmmode \lambda^{\prime} \else $\lambda^{\prime}$  \fi} 
\def \Vmax {\ifmmode V_{\rm  max} \else  $V_{\rm max}$  \fi} 
\def \Mgas {\ifmmode M_{\rm gas} \else $M_{\rm gas}$ \fi} 
\def \Mcg {\ifmmode M_{\rm cg} \else $M_{\rm cg}$\fi} 
\def \Mhg {\ifmmode M_{\rm hg} \else $M_{\rm hg}$ \fi} 
\def \Mdisc {\ifmmode M_{\rm disc} \else $M_{\rm disc}$ \fi} 
\def \Md {\ifmmode M_{\rm d} \else $M_{\rm d}$ \fi} 
\def \Mdm {\ifmmode M_{\rm DM} \else $M_{\rm DM}$ \fi} 
\def \Mda {\ifmmode M_{\rm d,0\%} \else $M_{\rm d,0\%}$ \fi} 
\def \Mdb {\ifmmode M_{\rm d,20\%} \else $M_{\rm d,20\%}$ \fi} 
\def \Mdc {\ifmmode M_{\rm d,40\%} \else $M_{\rm d,40\%}$ \fi} 
\def \md {\ifmmode m_{\rm d} \else $m_{\rm d}$ \fi} 
\def \Mb {\ifmmode M_{\rm b} \else $M_{\rm b}$ \fi} 
\def \Mbh {\ifmmode M_{\rm b,pri} \else $M_{\rm b,pri}$ \fi} 
\def \Mbs {\ifmmode M_{\rm b,sat} \else $M_{\rm b,sat}$ \fi} 
\def \zo {\ifmmode z_{0} \else $z_{0}$ \fi} 
\def \rd {\ifmmode r_{\rm d} \else $r_{\rm d}$ \fi}
\def \rg {\ifmmode r_{\rm g} \else $r_{\rm g}$ \fi}
\def \rb {\ifmmode r_{\rm b} \else $r_{\rm b}$\fi}
\def \rs {\ifmmode r_{\rm s} \else $r_{\rm s}$\fi}
\def \rc {\ifmmode r_{\rm c} \else $r_{\rm c}$\fi}
\def \rvir {\ifmmode r_{\rm vir} \else $r_{\rm vir}$\fi}
\def \rbh {\ifmmode r_{\rm b,pri} \else $r_{\rm b,pri}$ \fi} 
\def \rbs {\ifmmode r_{\rm b,sat} \else $r_{\rm b,sat}$ \fi} 

\title[Dark MaGICC: Dark Energy and galaxy formation]{Dark MaGICC: the effect of Dark Energy on galaxy formation. Cosmology does matter.}
\author[C. Penzo et al.]{C. Penzo$^{1}$\thanks{penzo@mpia.de}, A.V. Macci\`o$^1$, L. Casarini$^2$, G. S. Stinson$^1$, 
J. Wadsley$^3$ \\
\\$^1$ Max-Planck-Institut f\"ur Astronomie, K\"onigstuhl 17, 69117 Heidelberg, Germany
\\$^2$ Departamento de Fisica, Universidade Federal do Espirito Santo, Av. Fernando Ferrari 514, 29075-910 Vitoria (ES), Brazil
\\$^3$ Department of Physics and Astronomy, McMaster University, Hamilton, Ontario, L8S 4M1, Canada\\ 
 }

\begin{document}
\pagerange{\pageref{firstpage}--\pageref{lastpage}} \pubyear{---}
\maketitle
\label{firstpage}
\begin{abstract}

We present the Dark MaGICC  project, which aims to  investigate the  effect of  Dark Energy (DE) 
modeling on galaxy formation via hydrodynamical cosmological simulations.
Dark MaGICC includes four dynamical Dark Energy scenarios with time varying 
equations of state, one with a self-interacting Ratra-Peebles model.
In each scenario we simulate three galaxies with high resolution using smoothed particle hydrodynamics (SPH).
The baryonic physics model is the same used in the Making Galaxies in a Cosmological Context (MaGICC) project, and we varied only the background cosmology.
We find that the Dark Energy parameterization has a surprisingly important impact
on galaxy evolution and on structural properties
of galaxies at $z=0$, in striking contrast with predictions from  
pure Nbody simulations.
The different background evolutions can (depending on the behavior of the
DE equation of state) either enhance or quench star formation
with respect to a $\Lambda$CDM model, at a level similar
to the variation of the stellar feedback parameterization,
with strong effects on the final galaxy rotation curves.
While overall stellar feedback is still the driving
force in shaping galaxies, we show that the effect of the Dark Energy parameterization plays a larger role than previously thought, especially at lower redshifts. 
For this reason, the influence of Dark Energy parametrization on galaxy formation must be taken into account, especially in the era of precision cosmology.

\end{abstract}
\section{Introduction}
\label{intro}
Since   the   first   Type Ia   Supernova   data   were   published
(\citealt{Rie98},  \citealt{Per99}),  it  has   been  clear  that  our
Universe is expanding with a  positive acceleration. To enable an
accelerated expansion, there needs to be a repulsive force in our 
model of the Universe, and thus Einstein's Cosmological Constant
$\Lambda$ was  reintroduced. Its reintroduction lead to the current 
standard $\Lambda$ Cold Dark Matter ($\Lambda$CDM) cosmological 
model. 

Allowing the existence of a Cosmological Constant is the simplest solution
to obtain  a positive acceleration. It implies that more than 70\% of the 
energy in our Universe could be described as a
homogeneous fluid whose equation-of-state parameter is $w\equiv p/\rho
=  -1$   and, consequently, that  its   energy  density  $\Omega_\Lambda$
remains constant as a function of time. 

Despite  the excellent agreement of  $\Lambda$CDM cosmology with
observations, the model does suffer  from fundamental problems. 
The Cosmological Constant must be finely tuned in the early Universe
to reproduce the fit to observations we see today.
Moreover,  the attempt to explain the
presence of  such an energy density  with vacuum energy fails by several
orders of magnitude in predicting today's $\Lambda$ energy
density value. Finally it is a remarkable coincidence that the values of the
energy densities of $\Lambda$ and of matter are today of the same order
(see \citealt{Wei89},  \citealt{Car92}). 

For  these reasons cosmologists have been seeking alternatives to a
Cosmological  Constant. Such alternatives are generally  referred to 
as "dark  energy", a  more general  setting in
which the equation-of-state parameter  $w$ is allowed to be a function
of  time. Under  this assumption,  we describe dark energy with a  homogeneous
scalar field  whose energy  density evolves with  time. This changes the 
expansion history of the Universe and affects the evolution 
of density perturbations, \citet{Bal12}.
Thus, distinctive signatures  of dark energy models can be 
found by looking at the formation of structures.

In pioneering dark-matter-only simulations with an evolving $w$,
 \citet{Kly03} found that the differences between  the cosmological 
 models were not significant at z=0 both in the non-linear
matter power spectrum and in the halo mass function, although 
differences  between models  became significant  at higher
redshifts with a higher number of  clusters for the dark energy models
compared to  $\Lambda$CDM.

Subsequently, multiple groups investigated the properties of  dark matter
structures   in  DE   cosmologies (\citealt{Dol04}, \citealt{Bar05}  and \citealt{Gro09}).  They looked at halo concentrations,
velocity dispersions and abundance relations in dark energy
and  early   dark  energy   models.  More significant differences on the halo mass functions
between the  $\Lambda$CDM and dark energy cosmologies were found
at high redshifts (in these  works, all
models have the  same value for the mean  density amplitude $\sigma_8$
at z=0).

Several studies compared the inner structure of haloes simulated in $\Lambda$CDM
 and dark energy cosmologies in collisionless simulations.  In all cases, a Navarro-Frenk-White  
(\citealt{Nav97}) density profile well described the matter distribution.  The only 
difference in the dark energy simulations was that the matter was more centrally
concentrated because the haloes had earlier formation times (\citealt{Kly03,Lin03,Kuh05}).

While these studies all considered dark energy cosmologies that featured earlier collapse
times than $\Lambda$CDM, it is also possible for dark energy cosmologies to form structure
later.  The equation-of-state parameter, $w(a)$ can ``cross over the Cosmological
Constant boundary from below''.  In other words, $w$ can evolve from $w<-1$ at high
redshift to $w>-1$ at $z=0$.  Such models have \emph{less} collapsed structure at high
redshift than $\Lambda$CDM.  \citet{Xia06} and \citet{Xia13} showed observational
constraints favor such models.

While it is useful to study collisionless simulations of dark energy cosmologies,
we can only directly observe baryons.  Even though they account for $\sim\frac{1}{5}$
of the mass density of dark matter in the Universe, baryons can have an impact
on the formation of small scale structures \citep{Whi76,Zha04,Puc05,Jin06,Rud08,
Cas11a,Deb11,Van11,Cas12,Fed12}.  So far, simulations including dark energy 
have focused on massive galaxy clusters since cosmology has the largest effect 
on the formation of the largest structures.

In the last decade different groups have been studying
galaxy  formation   and  evolution   by  performing   high  resolution
hydrodynamical simulations in a cosmological context. Only recently
they have succeeded  in simulating realistic disk  galaxies, e.g. star
formation  history   matching  with  observational   constrains,  flat
rotation   curves,   exponential   surface   density   profiles   (see
\citealt{Rob06},  \citealt{Gov07},  \citealt{Age11},  \citealt{Gue11},
\citealt{Bro12},  \citealt{Sca12}, \citealt{Sti13},  \citealt{Mar13}). In all of these
 high  resolution simulations a \LCDM cosmology has
always  been assumed.
Recently an attempt to study galaxy formation in different
cosmological models has been presented in \citet{Fon12,Fon13}, where Nbody
simulations where combined with a Semi Analytical Model (SAM)
for galaxy formation. While they were able to address
the effect of cosmology on global properties of galaxies
(e.g. the cosmic star formation), due to their approach 
they were not able to study the effects of Dark Energy
parametrization on the internal structure of simulated 
galaxies.

In this work we aim to perform the first detailed
study of the effect of dark energy on galactic scale using
high resolution hydrodynamical simulations.
Our study is an extension of the MaGICC project  (Making
Galaxies In  a Cosmological  Context) and we dubbed it 
DarkMaGICC. The MaGICC project has been quite successful
in reproducing several properties of observed galaxies, including
star formation rates and stellar masses (\citealt{Bro12,Sti13}),
metals production and distribution (\citealt{Sti12,Bro13}), 
flat rotation curves and cored profiles (\citealt{Mac12,Dic14})
 and disc properties as observed in the Milky-Way
(\citealt{Bro13b,Sti13b}).

We adopt the same set of numerical parameters
describing the baryonic physics as in \citet{Sti13}, and 
perform high resolution hydrodynamical simulations with different
dark energy backgrounds, to study the impact
of Cosmology on galaxy properties.
In the spirit this paper is very similar to the recent 
work by \citet{Her14} that extended the MaGICC project
to Warm Dark Matter cosmologies.

This paper is organized as follows. In Section~\ref{models} the cosmological models
are described and compared with observational constraints. In Section~\ref{setup} we introduce the
numerical methods and implementations. In Section~\ref{results} we outline the results from our
set of simulations and investigate the interplay between feedback and dark energy. Finally we draw conclusions
in Section~\ref{conclusions}.

\section{Cosmological Models}
\label{models}
For  our project  we  have  chosen four  dynamical  Dark Energy  (dDE)
models, each of which is  consistent with WMAP7 data (\citealt{Kom11})
at the two sigma level. 
All the models have at $z=0$: $\Omega_{b_0} = 0.0458$, $\Omega_{DM_0} = 0.229$,
$H_0 = 70.2$ km$^{-1}$ s$^{-1}$ Mpc$^{-1}$, $\sigma_8$= 0.816, $n_s = 0.968$,
where these parameters are density parameters for baryons and dark matter,
Hubble constant, root mean square of the fluctuation amplitudes and
primeval spectra index.

\begin{figure}   
\psfig{file=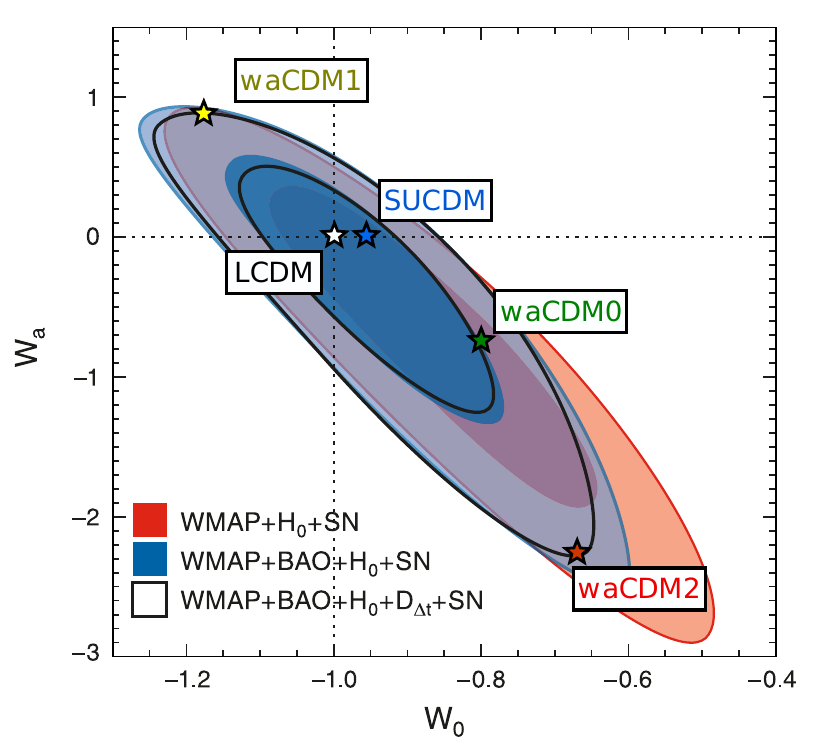,width=0.5\textwidth}
\caption{Confidence contours  of constrains  for $w_0$ and  $w_a$ from
  WMAP7. Each  cosmological   model   is   represented  by   a
  star. \textit{All}  of these  models are  \textit{viable} models
  according to WMAP7  data. The star representing  the SUCDM model
  is here shown only for comparison,  but clearly its position on this
  plot  holds only  at $z=0$,  since its  equation-of-state parameter
  $w(a)$ cannot be described by the CPL parametrization.}
\label{fig:wmap7}
\end{figure}

Three of the cosmological models, waCDM0, waCDM1 and waCDM2, are based
on a linear CPL parametrization (\citealt{Che01} and \citealt{Lin03b})
of the equation-of-state parameter $w$
\begin{align}
w(a) \equiv p(a) / \rho(a) = w_0 +(1-a) w_a
\end{align}
In Table~\ref{tab:param}  we show  the values we  chose for  $w_0$ and
$w_a$ in each of the three cases. waCDM0 is a model very close to LCDM
as shown in \citet{Cas10}, while 
waCDM1 and waCDM2, already studied in \citet{Cas11b}, are most 
distant in terms of w$_a$ on the w$_0$-w$_a$ contour plot in Figure~\ref{fig:wmap7}.

\begin{table}
\centering
\caption{Parameters of the waCDM cosmological models}
 \label{tab:param}
 \begin{tabular}{lll}
  \hline &  $w_0$ & $w_a$\\ \hline  waCDM0 (green) & -0.8 &  -0.755\\ waCDM1 (yellow) &
  -1.18 & 0.89\\ waCDM2 (red) & -0.67 & -2.28\\ \hline \hline
 \end{tabular}
\end{table}

We  have then  also included  a  fourth cosmological  model, which  we
called SUCDM, in which dark energy is described by a scalar field with
a SUGRA (SUper GRAvity) self-interacting potential of the form
\begin{align}
V(\phi) = \frac{\Lambda^{4+\alpha}}{\phi^{\alpha}}\exp(4\pi G \phi^2)
\end{align}
where we chose $\alpha= 2.9$ and $\Lambda=10$ GeV in agreement with \cite{Ali10}.

In  Figure~\ref{fig:wmap7} the  two-sigma contours  from WMAP7  in the
$w_a$-$w_0$  plane   are  shown,   and  each  cosmological   model  is
represented by a  triangle. It is important to  note that \textit{all}
of  these  models  are   \textit{viable}  models  according  to  WMAP7
data. The triangle representing the SUCDM model is here shown only for
comparison, but clearly its position on this plot holds only at $z=0$,
since it's  equation-of-state parameter $w(a)$ cannot  be described by
the CPL parametrization.\\

\begin{figure}   
\psfig{file=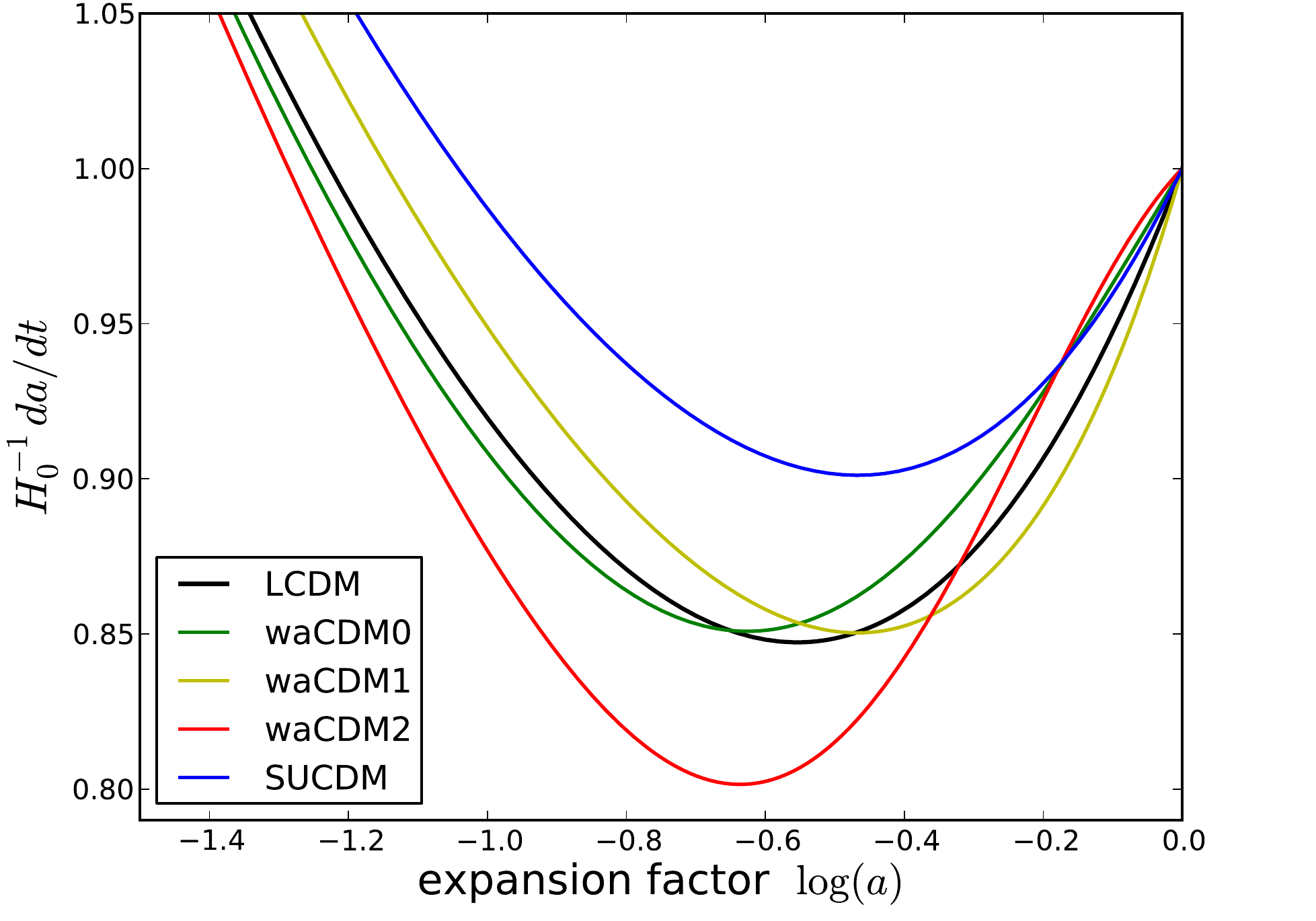,width=0.5\textwidth}
\caption{Expansion velocities of  the universe in units  of the Hubble
  constant  as  a  function  of  the scale  factor. Different colors represent
  different cosmological models.}
\label{fig:dadt}
\end{figure}

In  order to  show how  the  background evolution  of these  different
cosmological  models changes,  in  Figure~\ref{fig:dadt}  we show  the
expansion  velocity  of  the  universe in  all  the  different  chosen
cosmologies.  We chose  to  compare different  cosmological models  by
normalizing them  to the  same $\sigma_8$ today.  With this  choice, a
model with a  faster expansion will have to  start producing structure
earlier than a  model with slower universe  expansion. This
means that statistically,  the SUCDM model (blue)  will show collapsed
structures  at an  earlier epoch  than the  waCDM2 model  (red), in
order to  compensate for  the faster expansion  of the  universe. The earlier
structure formation also leads to earlier accretion of the substructures
onto the main halo. 
In turn, we expect that earlier accretion will lead to earlier star formation
in  the  simulated galaxies.

\section{Numerical methods}
\label{setup}
\subsection{Initial Conditions and Dynamical Dark Energy generalization}
We   modified  the   initial   condition   generator  {\sc   grafic-2}
(\citealt{Ber01}) such that  we can generate initial  conditions for a
generic  cosmological model  once  the evolution  of the  cosmological
parameters  are  given  as  an  input.  Our  implementation  requires
transfer  functions for  baryons and  for dark  matter at  initial and
final redshift,  evolutions of  the density parameters  $\Omega_i$,
linear growth factor D$_{+}$ and f$_{\Omega}$, the logarithmic derivative 
of the growth factor with respect to the scale factor.

We  have computed the transfer functions
with a modified version of {\sc camb} (\citealt{Lew02}) that allows us
to account for dynamical dark  energy scenarios. As the original code,
{\sc grafic-de} is  able to generate zoomed initial  conditions from a
cosmological box.

At first we generate a uniform particle distribution in a 80 Mpc/h box
with $350^3$ particles. The initial conditions were evolved
with the {\sc pkdgrav} tree-code (\citealt{Sta01}), we then select a dark matter
halo and we re-simulate it with {\sc pkdgrav} for the dark-matter-only
runs  and with  {\sc  gasoline} \citet{Wad04}  for the  hydrodynamical
runs. The $\Lambda$CDM halos are chosen so that no structures are present in within
three of their virial radii and the equivalent halos in the other cosmological
models are then identified.
 
Both {\sc pkdgrav} and {\sc gasoline} have been modified so that
they can  accept cosmological background  evolutions as  inputs. We  have simulated
three  different galaxies,  gal$\alpha$, gal$\beta$,  gal$\gamma$, and
each  of  them was  then  run  in  all  five cosmological  models.  In
Table~\ref{tab:param2} we  summarize the main proprieties  of the three
galaxies in \LCDM cosmology that we have chosen for this project.
For all three galaxies the softening for gas and dark matter particles are
respectively 0.45 and 1 kpc. Note that, for all  galaxies, the SUGRA equivalents are always the most
massive and, on the other hand, the waCDM2 are always the least massive.

\begin{table}
\centering
  \caption{Physical  properties  of  the  selected  galaxies  for  the
    respective  \LCDM cases. We show virial radius, virial mass (total mass), dark matter mass,
    gaseous mass and stellar mass, respectively calculated within one virial radius.}
 \begin{tabular}{@{}cccccccr@{}}
  \hline &  \Rvir &  \Mvir &  \Mdm &  \Mgas &  $M_ {*}$  \\ &  [kpc] &
         [\Msun] & [\Msun] & [\Msun] & [\Msun] \\ \hline gal$\alpha$ &
         240 & $7.7\times 10^{11}$  & $6.8\times 10^{11}$ & $5.3\times
         10^{10}$  &  $4.5\times  10^{10}$   \\  gal$\beta$  &  227  &
         $6.6\times  10^{11}$  &   $5.8\times  10^{11}$  &  $4.4\times
         10^{10}$  &  $3.9\times  10^{10}$  \\  gal$\gamma$  &  184  &
         $3.4\times  10^{11}$  &   $3.1\times  10^{11}$  &  $2.7\times
         10^{10}$ & $6.6\times 10^{9}$ \\ \hline \hline
          \label{tab:param2}
 \end{tabular}
\end{table} 

\subsection{Hydrodynamical Simulation} 
\label{hydro}

For  the  DarkMaGICC project  we  are  using the  same  baryonic physics that was
used in the MaGICC  project (see \citealt{Sti13}),
based on  the  smoothed  particle  hydrodynamics  (SPH)  code
\textsc{gasoline}  (\citealt{Wad04}).  For   further  details  on  the
physical processes  implemented in  \textsc{gasoline} please  refer to
\citet{Sti13}.  Briefly,  stars form  from  cool  dense gas  that  has
reached a temperature of  $T = 1.5\times 10^4$ K and  a density of 9.6
cm$^{-3}$ following  the Kennicutt-Schmidt Law with 10\% efficiency 
of turning gas into stars during one dynamical time (\citealt{Sti06}). 
The stellar mass distribution in each star particle follows the 
Chabrier initial mass function (IMF), \citet{Cha03}.

Massive stars explode as type II supernovae and deposit an energy of $E_{SN}=10^{51}$ ergs 
into the surrounding gas. Cooling for gas particles subject to supernova feedback is
delayed based on the sub-grid approximation of a blast wave as described in \citet{Sti06}.

Furthermore, radiation energy from massive  stars is  considered  
since molecular  clouds are 
disrupted  before the first  supernova  explosion  (which
happens  after 4  Myr from  the formation  of the  stellar
population). We assume that 10\% of the total radiation energy 
is coupled with the surrounding gas.
The inclusion of this early stellar feedback reduces star formation before supernovae start
exploding. Thus, after  the early stellar feedback heats the  gas to T
$>10^6$ K,  the gas rapidly cools  to $10^4$ K, which  creates a lower
density medium than if the gas  were allowed to continue cooling until
supernovae  exploded.
\citet{Sti13} shows how  this feedback  mechanism
limits star formation to the amount prescribed by the stellar-halo mass
relationship at all redshifts.
The code also includes metal cooling and metals can 
diffuse between gas particles as described in Shen \etal (2010).

The hydrodynamical simulations
in \emph{all} cosmological  models have been run  with the \emph{same}
feedback descriptions  just mentioned. The  main point of the  work is
not which of the many recently used feedback recipes best
reproduced  observations,  but  the  impact  of
dynamical Dark Energy on galactic scales.
\section{Results}
Using hydrodynamical and dark matter only simulations, we present how gal$\alpha$, 
gal$\beta$ and gal$\gamma$ evolved and their $z=0$ properties.  
These include the dark matter distribution, gas, star and total halo masses, 
star formation histories, baryonic matter distribution (rotation curves and surface 
brightness profiles), and the chemical enrichment of the galaxies. 
\subsection{Stellar and Halo Mass}
\begin{figure}   
\psfig{file=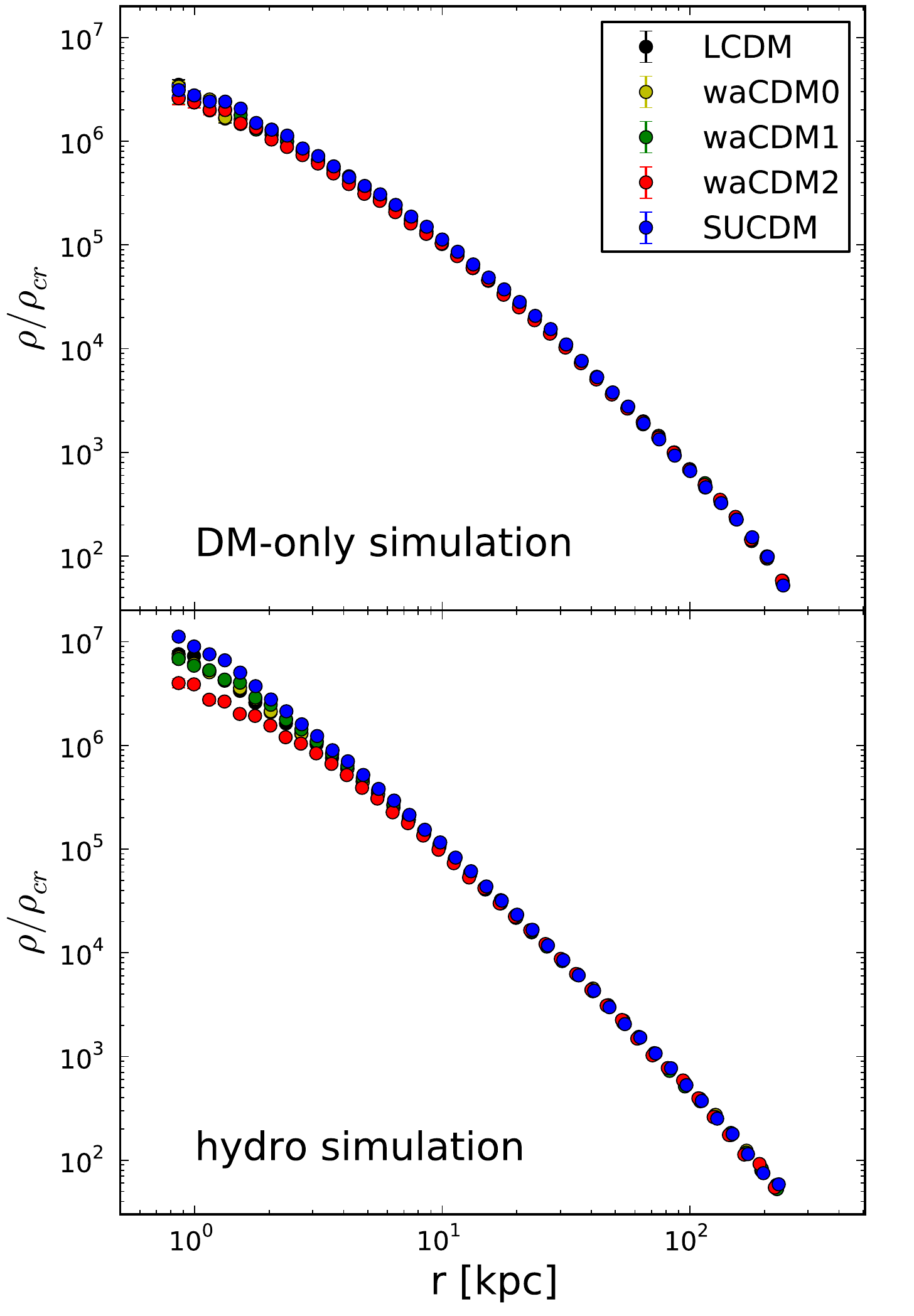,width=0.525\textwidth}
\caption{Radial  density  profile  of  gal$\alpha$ simulated in  all  five
  cosmological  models,  respectively in  a  dark  matter only  (upper
  panel) and in a hydrodynamical simulation (lower panel). We plot the
  density in units of critical density, $\rho_{crit} = \frac{3H_0^2}{8
    \pi  G}$ with  G  gravitational  constant, as  a  function of  the
  distance from the center of mass of the galaxy. Different colors represent 
  different cosmological models.}
\label{fig:density}
\end{figure}

Fig.~\ref{fig:density} shows how the dark matter profiles of simulations with and without
baryons compare in gal$\alpha$ for all different cosmological models.  
Gal$\beta$ and gal$\gamma$ show similar results.
\label{results}
The four radial density profiles from the dark matter only simulations (top panel)
are almost indistinguishable. This confirms previous findings from N-body simulations 
(see Section \ref{intro}), i.e. that dark matter only simulations on galactic scales weakly 
depend on the dark energy model.

The lower panel of Fig.~\ref{fig:density} shows the radial density profiles of dark
matter in hydrodynamical simulations. In contrast to the dark matter only simulations,
 the density profiles vary depending on the dark energy model used. 

\begin{figure}   
\psfig{file=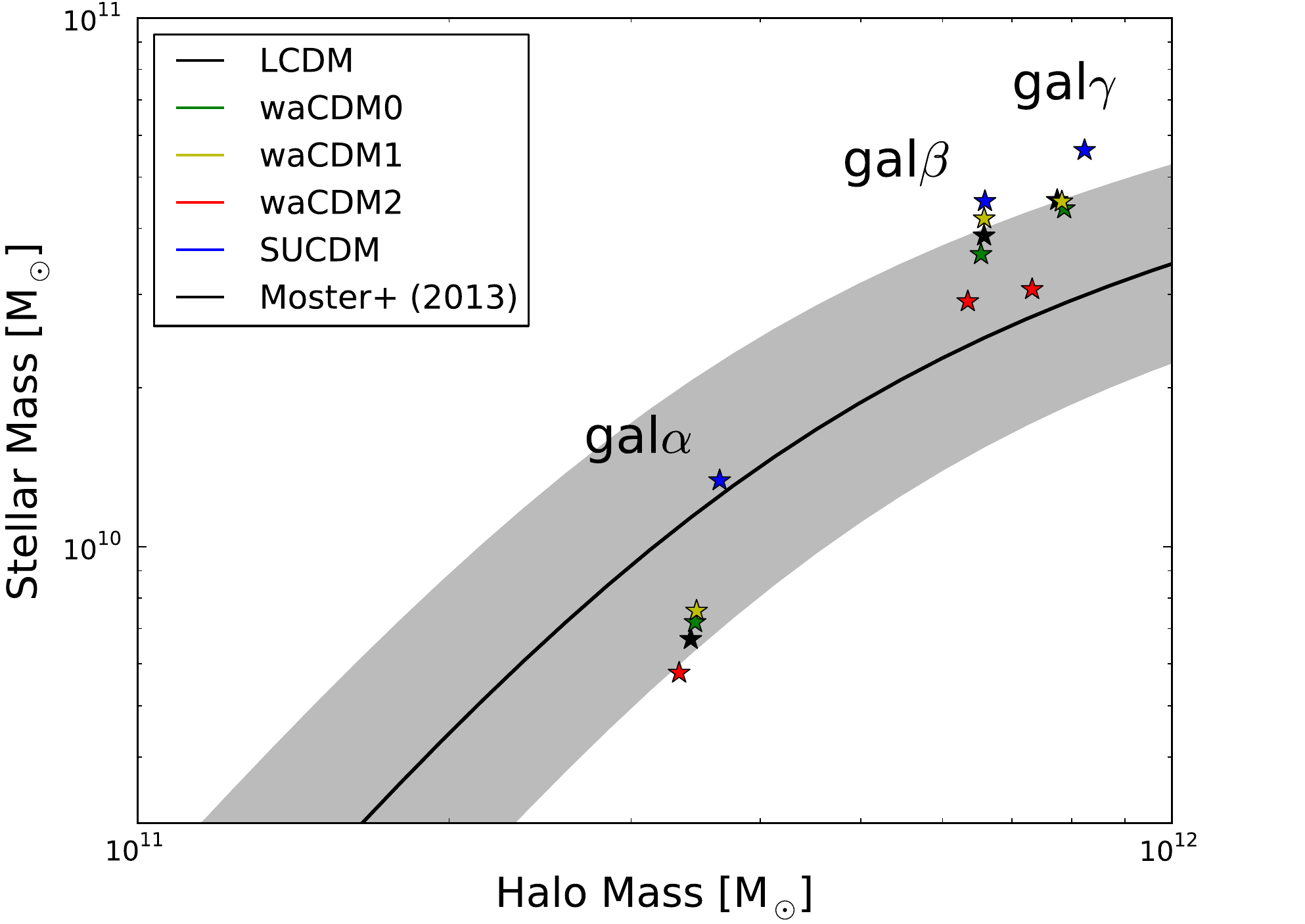,width=0.525\textwidth}
\caption{Stellar mass  as a  function of  halo mass  at $z=0$  for
 gal$\alpha$, gal$\beta$, gal$\gamma$ simulated in all the different cosmologies.}
\label{fig:moster_z0}
\end{figure}

Fig.~\ref{fig:moster_z0} sets gal$\alpha$, gal$\beta$ and gal$\gamma$ in the
abundance matching plot at $z=0$. The  black line represents
the prediction obtained by the abundance matching technique (see \citealt{Mos13}),
and  the shaded  area represents  the  errors on  the prediction.  The abundance matching 
prediction does not  vary from \LCDM to the  other cosmologies since all cosmological
models have the same value  for  $\sigma_8$  at  $z=0$.
While statistical conclusions are not possible because of the limited sample,
the three galaxies show the same trend 
as  a function of  cosmology. By simply varying the cosmological model, the change 
among the three galaxies is of only a few percent in the dark matter mass, 
while the stellar mass almost doubles. Galaxies simulated in the waCDM2 
cosmology (red symbols) always make the least stars at $z=0$, while the
galaxies formed in the SUGRA cosmology (blue symbols) always make the most stars.
Galaxies formed  in a \LCDM cosmology always lie in the middle. 
The hierarchy is in agreement with the behaviors of the cosmological background evolutions
of these cosmological models (Fig.~\ref{fig:dadt} and Section~\ref{intro}), since
we expect more structures to be formed in a cosmological model that begins
forming structures earlier.
\subsection{Evolutions of the $M_\star-M_h$ relationship}
\label{mstarmhalo}
\begin{figure*}   
\centering
\begin{minipage}{180mm}
\psfig{file=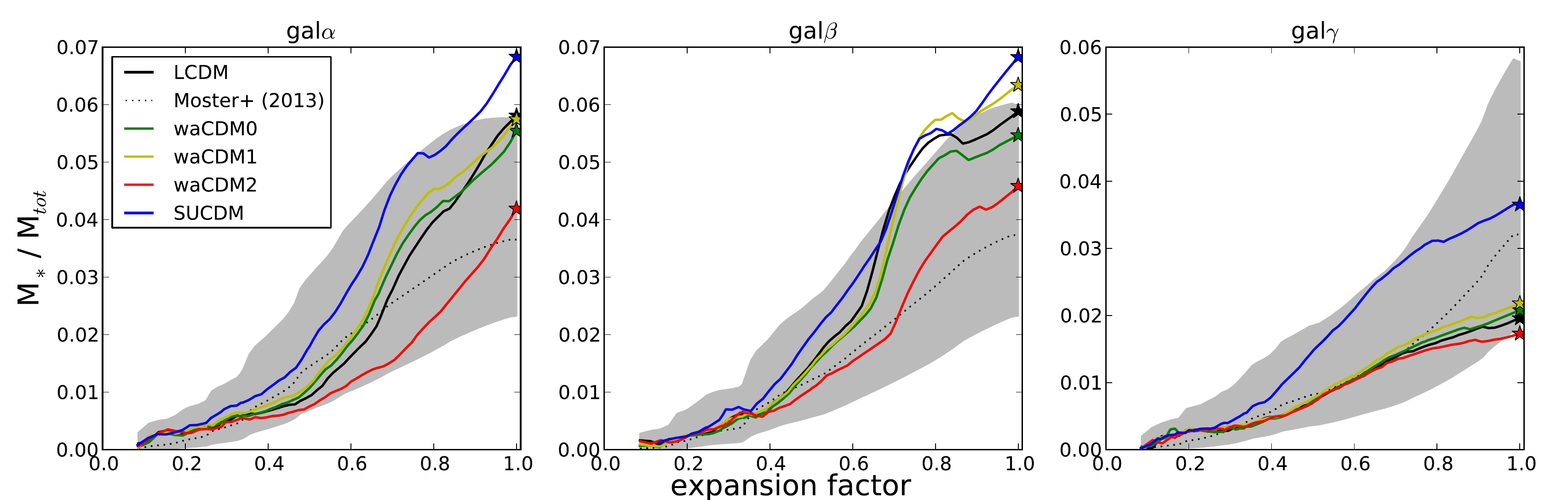,width=1.0\textwidth}
\caption{Evolutions of the stellar-halo mass relation as a function of
  expansion factor for gal$\alpha$, gal$\beta$ and gal$\gamma$.}
\label{fig:moster_z}
\end{minipage}
\end{figure*}

Fig.~\ref{fig:moster_z} shows how the ratio of stellar mass and
total mass  evolve with expansion factor  $a = 1 /  (z+1)$. Each panel
relates to  a specific galaxy  and the different colors  describe each
galaxy  run in  a  different  cosmology. Again, the black solid  line
represents  the  expected  evolution  for  a  \LCDM  model  using  the
abundance matching technique. The  predicted evolutions do change with
the  change in  cosmology, but  they  all do  not distance  themselves
significantly from the  \LCDM prediction. Hence, out of  clarity, we have
only  plotted the  \LCDM  predicted behavior  from abundance  matching.
As in the $z=0$ case, the $M_\star-M_h$ trends for the galaxies simulated
in different cosmologies are in agreement with the evolution of their cosmological backgrounds.
In the SUCDM cosmology, we expect higher density perturbations to compensate
for the faster expansion of the Universe. These higher density perturbations
trigger a more efficient star formation (blue lines in Fig.~\ref{fig:moster_z}).
On the contrary, the waCDM2 galaxy (red lines) always makes
less stars throughout its evolution.
The cosmological models waCDM0 and waCDM1 are not far apart from the
\LCDM model, in the $w_a - w_0$ plane, thus we would expect galaxies that 
live in those models not to differ greatly from galaxies that live in the \LCDM universe.
This expectation is nicely reproduced for all three haloes.  

\begin{figure*}  
\centering
\begin{minipage}{180mm}
\psfig{file=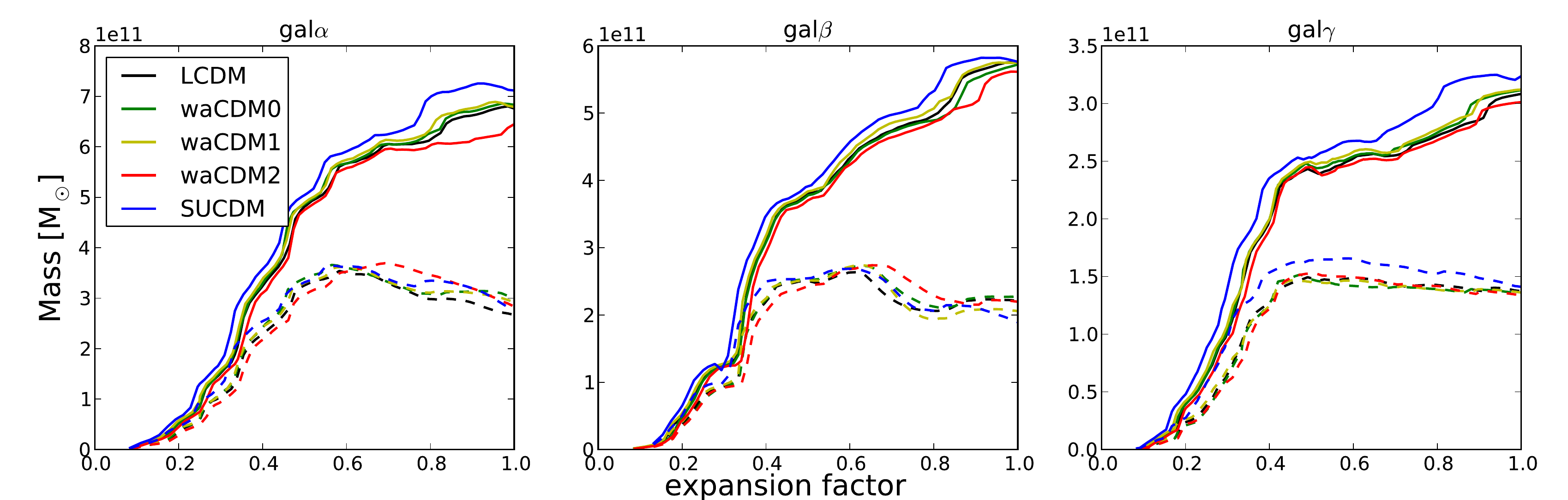,width=1.0\textwidth}
\caption{Evolution  of  the dark matter mass (solid  lines) and  gas  mass
  (dashed lines) as a function of scale factor for the for gal$\alpha$,
  gal$\beta$ and gal$\gamma$ in all different cosmological models. For
  an easier comparison,  the gas mass was increased of a factor
  of five.}
\label{fig:vir}
\end{minipage}
\end{figure*}

As shown in Fig.~\ref{fig:moster_z} it is noticeable how both gal$\alpha$ and gal$\beta$ undergo
a significant merger  around $a=0.8$ which raises their star formation
efficiency and increases their dark matter mass. The merger is visible also from the dark matter mass
of the halos as a function of the scale factor. Fig.~\ref{fig:vir} shows a clear increase
in the dark matter mass due  to the accretion of a nearby satellite galaxy.
\subsection{Star Formation Histories}
\label{sfh}
\begin{figure*}   
\begin{minipage}{180mm}
\psfig{file=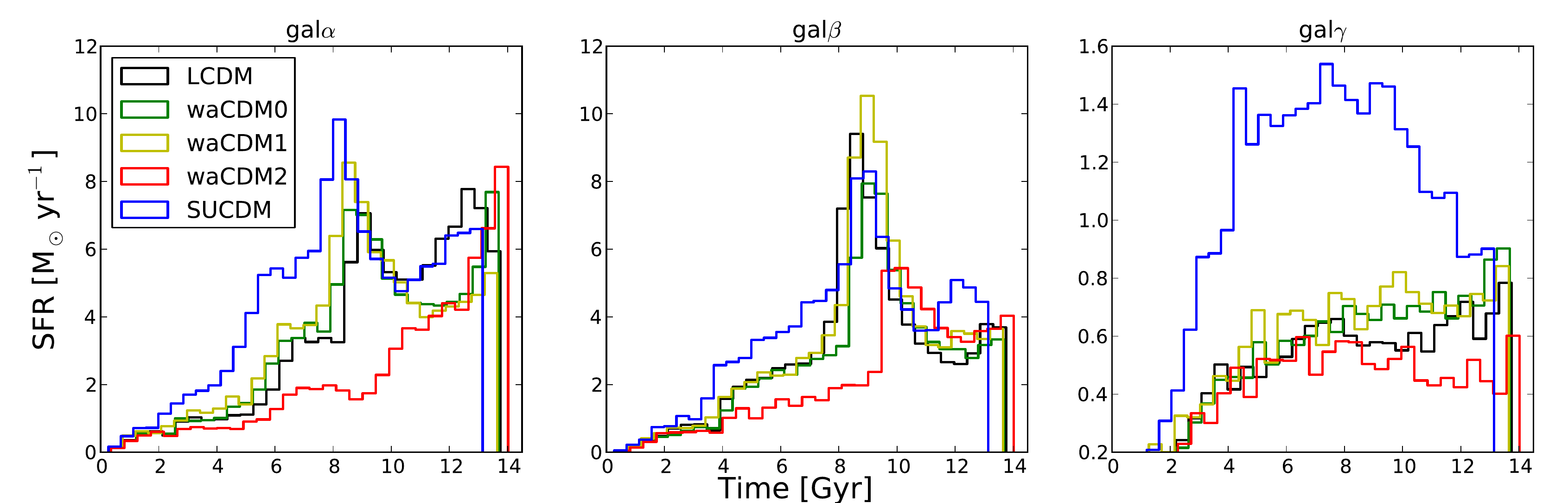,width=1.0\textwidth}
\caption{Star  formation  histories  for gal$\alpha$,  gal$\beta$  and
  gal$\gamma$ in all the cosmological models.}
\label{fig:sfh}
\end{minipage}
\end{figure*}

Fig.~\ref{fig:sfh} shows  the star formation  rate (SFR)  as a
function of physical time. At $z=0$ the different cosmological
models show  longer or  shorter ages  of the  Universe because 
how much physical time elapses as the Universe expands depends 
on the choice of cosmology. The choice  of showing the
star formation in standard physical  units gives more insight. 

Fig.~\ref{fig:sfh} shows how dark energy can suppress  and delay star formation.
Interestingly, the  waCDM2 cosmology
(red lines) \emph{delays} star  formation, both  in the
case  of gal$\alpha$  and gal$\beta$. In  all three  galaxies the
waCDM2  cosmology drastically \emph{suppresses}  star
formation until recent times.

As  pointed  out  in Section~\ref{mstarmhalo},  both  gal$\alpha$  and
gal$\beta$ undergo  a merger. The merger event is clearly marked by the presence of a
peak in the SFRs between 7 and 10 Gyrs (notice how the peak shifts in time according
to the cosmological model). After the  star formation burst due to the
merger, both  galaxies decrease their  star formation activity  due to
the decrease in the amount of available cold gas. This is shown in Fig.~\ref{fig:vir},
 where  we plot the  evolution of the dark matter mass
and the cool gas mass (T$<10^5$ K).
\subsection{Rotation Curves}
\begin{figure*}   
\begin{minipage}{180mm}
\psfig{file=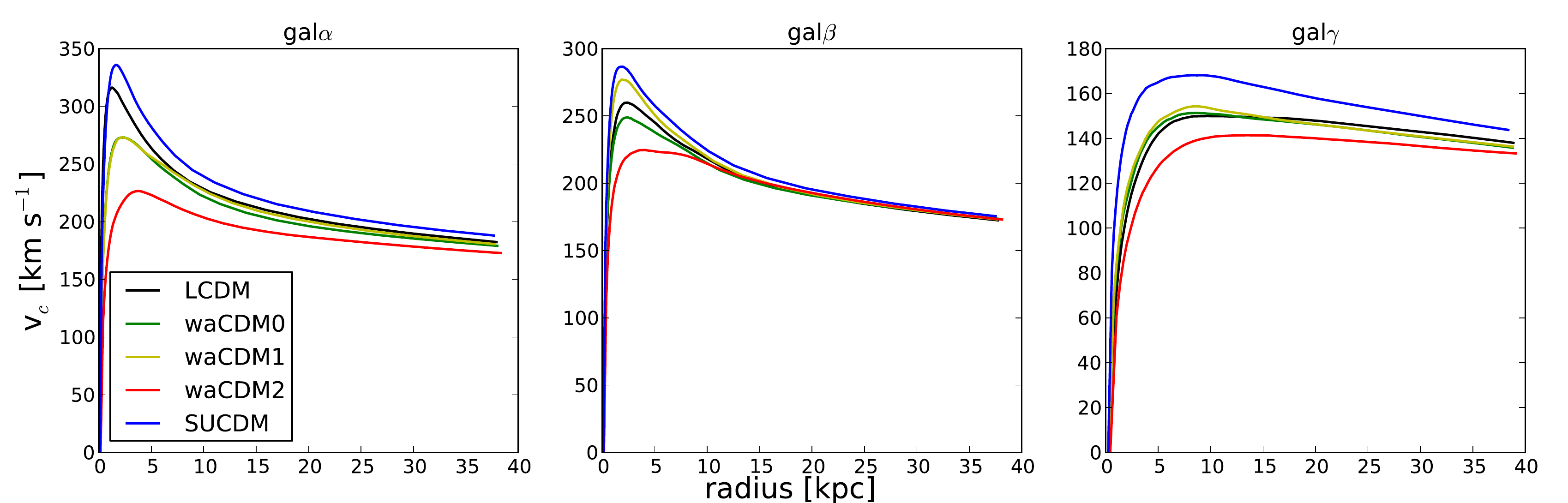,width=1.0\textwidth}
\caption{Rotation curves  for gal$\alpha$, gal$\beta$  and gal$\gamma$
  in all the cosmological models.}
\label{fig:rcs}
\end{minipage}
\end{figure*}  
Different star formation histories reflect different matter distributions 
among the galaxy, as the rotation curves in Fig.~\ref{fig:rcs} show.
Galaxies with delayed star formation (waCDM2  
cosmology, red lines in Fig.~\ref{fig:sfh}) have flatter rotation 
curves than galaxies where star formation started earlier (SUCDM 
cosmology, blue lines), see \citealt{Sti13}.  
Thus, a galaxy can have a flat or centrally peaked rotation curve based
simply on the background cosmology in which it forms.

Centrally peaked rotation curves have long been the prime symptom 
of the overcooling problem in disk galaxy formation simulations, see \citet{Sca12}. 
In the centers of halos the gas density becomes high enough that hot gas starts radiating
and consequently cools. In such environments, the cooling process is unstable because, 
once the hot gas has cooled, it no longer pressure supports the surrounding gas, which 
then becomes denser and cools. Stars then form in excess and primarily in the central 
concentration, and they produce peaked rotation curves.

Most solutions have focused on adding energy from stars or AGN 
(\citealt{Sca12}).  \citet{Sti13} showed one solution based on stellar 
winds from massive stars (i.e. ``early stellar feedback''). Our results show 
that also cosmology can have a considerable effect on flattening rotation curves.

This work shows that \emph{simply} changing  the evolutions of the 
dark energy equation of state \emph{flattens} rotation  curves of a considerable
and definitely  \emph{observable} amount (i.e.  more than 100  km/s in
both  gal$\alpha$   and  gal$\beta$).
\begin{figure}   
\psfig{file=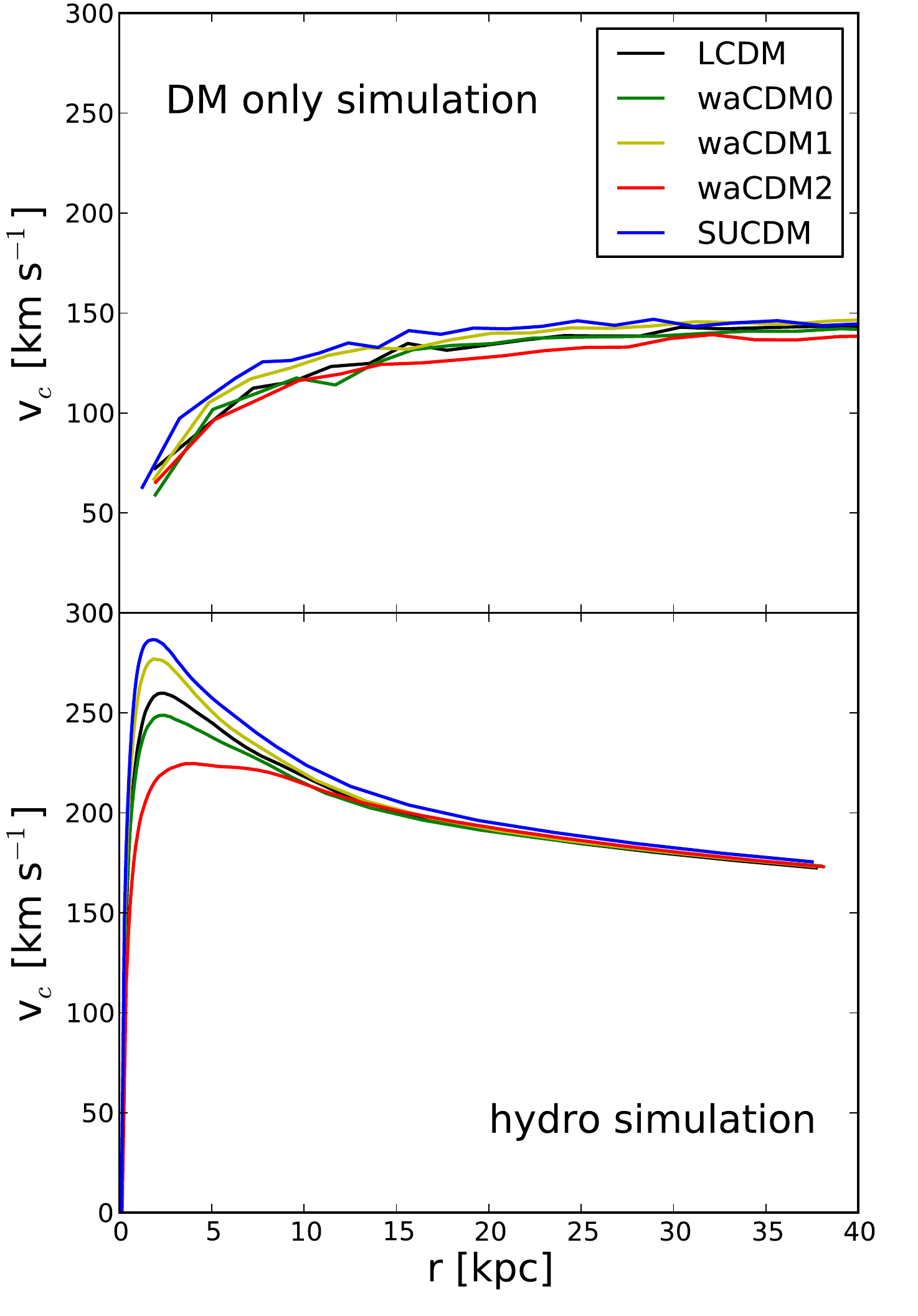,width=0.525\textwidth}
\caption{Rotation curve for gal$\alpha$  in all different cosmological
  models, respectively in a dark  matter only simulation (upper panel)
  and in a hydrodynamical simulation (lower panel).}
\label{fig:dm}
\end{figure}  
Fig.~\ref{fig:dm} compares rotation curves for  gal$\alpha$ in dark  matter  only  simulations
(upper  panel)  and  in  SPH simulations  (lower   panel) for each cosmological
model. The  change is striking. While in the dark matter only case the cosmological models
are almost indistinguishable, they become clearly distinguishable in the 
hydrodynamical simulations.

\citet{Sti13} show that early stellar feedback is a key ingredient to simulate
realistic disc galaxies. In particular, early stellar feedback can flatten rotation curves.
Unexpectedly, at $z=0$ the effect of early stellar feedback is comparable with the effect of
dynamical dark energy.
\subsection{Disc Scale Lengths}
\begin{figure*}
\centering
\begin{minipage}{180mm}
\psfig{file=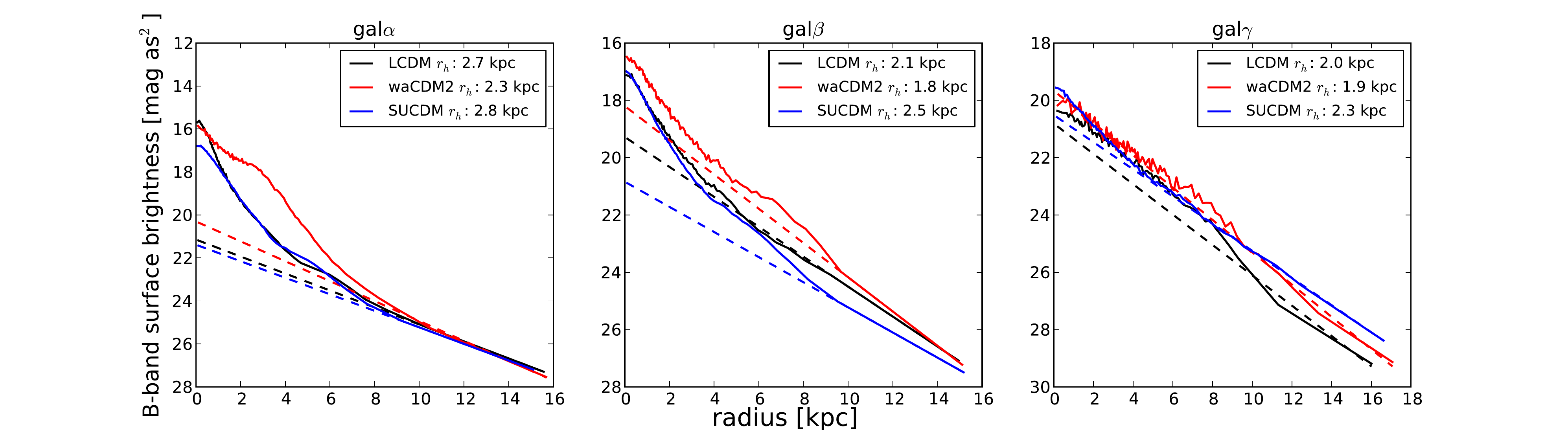,width=1.0\textwidth}
\caption{Disc scale lengths for gal$\alpha$, gal$\beta$  and gal$\gamma$ for \LCDM, waCDM2 (red lines) and SUGRA (blue lines). For clarity, only these two models are shown.}
\label{fig:discs}
\end{minipage}
\end{figure*} 
Differences due to cosmologies are also visible in the surface brightness profiles at z=0, as Fig.~\ref{fig:discs} shows. Here we plotted only the two extreme cosmological models out of clarity.
As seen in Section~\ref{sfh}, waCDM2 cosmology (red lines) is able to suppress and delay star formation until low redshifts, when star formation finally starts to increase.
In turn, a suppressed and delayed star formation affects the shapes of the surface brightness profiles, especially in the two most massive galaxies. Gal$\alpha$ and gal$\beta$ show steeper profiles in their inner regions (for further discussion on effects of delayed star formation on surface brightness profiles, see \citealt{Sti13b}). 

The rest of the cosmological models do not significantly affect the profiles nor the scale lengths. As shown in the previous sections, the effect of cosmology seems to increase when the galaxy has a mass that is close to the peak of baryonic efficiency, and this is confirmed also when looking at the surface brightness profiles, where the effects of cosmology on gal$\gamma$ are not significant.
\subsection{Feedback and cosmology, metallicity interplay}
We showed that a  model whose universe velocity expansion is
slower compared to the one of  \LCDM (e.g. waCDM2, red lines) has a
lower star  formation till much later  times, and on the  other hand a
model  whose   universe  velocity  expansion  is   faster  than  \LCDM
(e.g. SUGRA, blue lines) has a  higher star formation at all redshifts
(see Figure~\ref{fig:sfh}).  We can trace  back  this difference  to  the  
fact  that all  five  different
cosmological  models  have  the  same  $\sigma_8$  today,
because, in  order for  this to  happen, structures  in a  SUGRA model
(blue lines) have  to start forming earlier. This implies  that, at the
starting   redshift  ($z=99$   for   all   simulations),  the   density
perturbations that seeded structure formation had to be slightly bigger in
the  SUGRA  model  (blue  lines)   compared  to  the  initial  density
perturbations in  the waCDM2  model (red lines). Thus, stars will 
start forming earlier since more gas is accreted and cools.  

\begin{figure}   
\psfig{file=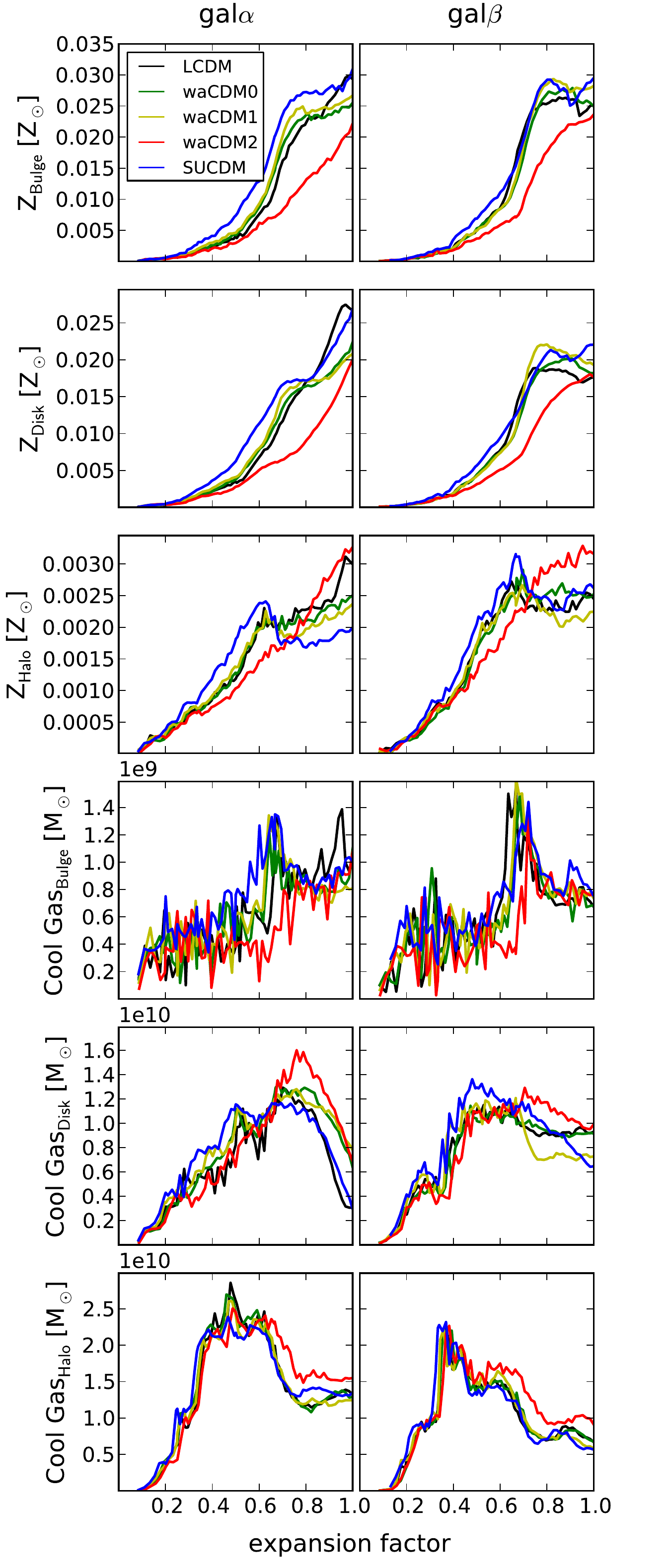, width=0.55\textwidth}
\caption{Mean metallicity in solar units and cool gas in solar masses (T $<10^5$K) for gal$\alpha$ and gal$\beta$ as function of scale factor in ``bulge'', ``disc'' and ``halo''. Different colors represent different cosmological models.}
\label{fig:metalsandgas}
\end{figure}

These differences in the initial perturbations do not significantly affect properties
of structures on galactic scales in dark matter only simulations.
On the other hand, the  interplay between  cooling, metallicity  and
star formation not  only  helps differentiating between the cosmological  
models but also enhances their differences. To highlight the Òpositive feedbackÓ 
star formation has on radiative cooling through metal enrichment, the first three rows of Fig.~\ref{fig:metalsandgas} show
the evolution of metallicity as a function of scale 
factor for three different regions of gal$\alpha$ and gal$\beta$, a central 2 kpc 
sphere (``bulge''), a disk cylinder with radius 20 kpc and 6 kpc thickness
 (``disc''), and a sphere of the size of the $R_{vir}$ (``halo'').

The waCDM2 model (red) exhibits the lowest metallicity in the bulge and disc
throughout its evolution, which reflects its lower star formation rate and hence 
lower enrichment rate.  The effect of increased metal enrichment is non-linear:  the more star 
formation enriches gas, the faster the gas cools, and the more stars that 
subsequently form. 

The halo metallicity of waCDM2 is also lower throughout 
most of the galaxy evolution, but becomes higher after $a\sim0.75$, as its mean
halo metallicity continues increasing while the metallicity in the other models starts
 to decrease or flatten out at that time.  
Both waCDM2 galaxies start to have  more metallicity in the halo  due to Supernova
explosions being able  to move the gas outside from  the disc. 

Comparing the
trends  for  metallicity,  cool  gas  (Fig.~\ref{fig:metalsandgas}) and star formation rates (Fig.~\ref{fig:sfh})
as functions  of scale  factor, we  find them  in agreement.  Because of the lower metallicity,
the waCDM2 model (red line) ends
up having the least amount of gas  that has been able to cool and thus
also  makes the  least  amount of  stars. Having  used up  a
smaller amount  of cold gas at  earlier times increases the  amount of
cold gas left for star formation at late times. The presence of cool gas
that has yet not formed stars can be seen in Fig.~\ref{fig:metalsandgas},
where after $a=0.7$ the disc of  the waCDM2 galaxy has the most amount
of  cool  gas compared  to  the  galaxies  in the  other  cosmological
models.  This is  also  the case  for  the galaxy  halo,  and this  is
probably due to the cooled gas moved by supernova explosions.
\subsection{Feedback--Dark Energy degeneracy}
\begin{figure*}   
\begin{minipage}{180mm}
\psfig{file=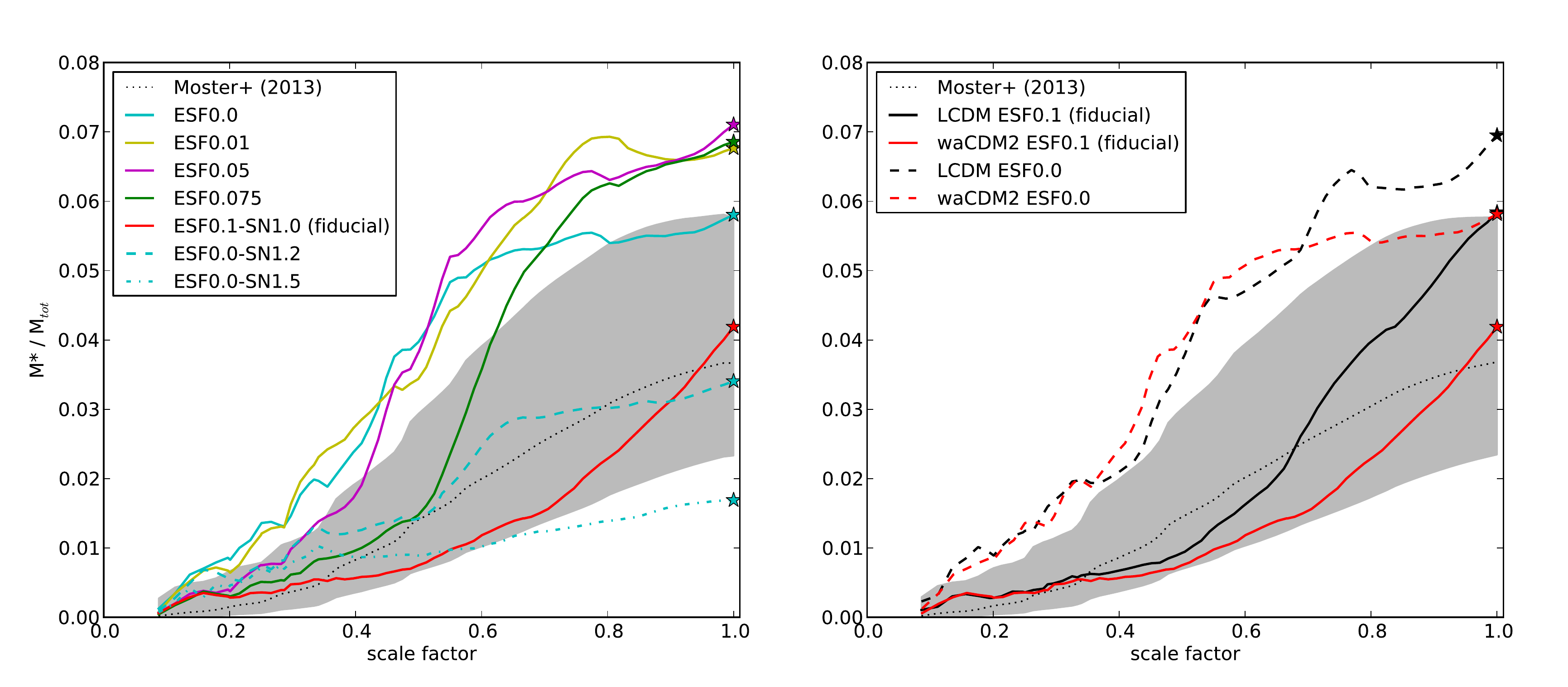,width=1.0\textwidth}
\caption{Evolution of stellar-halo mass relation with scale factor. We
  are now showing  only the case of gal$\alpha$. In  the left panel we
  have changed  the feedback  parameters for the  waCDM2 run.  We have
  increased  the early  stellar feedback  parameter from  zero to  the
  fiducial value  (see \citet{Sti13})  while keeping the  SN parameter
  fixed  to the  fiducial value  of 1.0  (solid lines).  We then  have
  changed the  SN parameter while  keeping the early  stellar feedback
  fixed to zero (dashed and  dash-dotted cyan lines). The dotted black
  line is the  abundance matching prediction and the  shaded area its
  errors. In  the right panel we  compare the effect of  early stellar
  feedback feedback with the effect of different cosmology, \LCDM (red
  and black lines) and waCDM2 (blue and red) }
\label{fig:feedback}
\end{minipage}
\end{figure*} 
Along with dark energy having a profound effect on galaxy evolution, 
galaxy  formation strongly  depends  on  the feedback modeling.  
\citet{Sti13} showed that pre-supernova feedback primarily suppresses 
star formation at early stages of evolution, which is similar to the evolution 
seen in the dark energy models that have the most delayed expansion.
Thus, we wish to explore whether dark energy or 
stellar feedback has a greater effect at early times. 

We select the waCDM2 cosmology (red lines in previous plots), which showed the most star
formation suppression and delay, and we re-simulate it with a range 
of stellar feedback strengths. We vary both the supernova feedback 
efficiency and the early stellar feedback separately.

First, the early stellar feedback is turned from 10\% down to 0\% efficiency 
with the standard $10^{51}$ erg of supernova energy.  Then, with 0 early
stellar feedback, the supernova feedback strength is increased to 120\% 
and 150\%.

The left panel of Fig.~\ref{fig:feedback} shows each of these variations 
implemented in the waCDM2 model for gal$\alpha$. The stellar mass evolution shows 
a strong dependence on the early stellar feedback parameter.  
A decrease of 25\% to 7.5\% increases the final stellar mass 50\% and moves most 
of the star formation from late to early times.
All the simulations with less than 7.5\% efficiency, but more than 0 early 
stellar feedback end with nearly the same final stellar mass.  What is 
somewhat surprising is that the simulation with 0 early stellar feedback 
ends with {\emph less} stellar mass than these intermediate feedback models. 
\citet{Sti13} also saw this effect 
and found that it was due to the higher star formation efficiency at early 
times driving stronger outflows due to the greater supernova feedback.  
Thus, gas was driven to radii where it could not be reaccreted, whereas 
the early stellar feedback does not drive gas so far away.

Supernova  feedback
unambiguously decreases the amount of stars formed throughout the
galaxy's evolution, but can easily push the trends out of the expected
behaviors from abundance matching techniques suggesting that Supernova
feedback alone is  not enough to reproduce realistic  galaxies. 

The right panel of Fig.~\ref{fig:feedback} shows a comparison of the 
stellar mass-halo mass evolution of the waCDM2 model (red lines) with 
the LCDM model (black lines) for gal$\alpha$. The models separate 
most notably at late times. The corresponding simulations with no early stellar
feedback are shown in the dashed lines. They clearly show that cosmology 
has the strongest effect at late times (i.e. after  $a=0.7$) and pre-SN feedback has the strongest effect at high redshift.

\section{Conclusions}
\label{conclusions}
The  intention of  this work  was to  investigate for the first time 
the  effect of  dark energy on galactic scales in SPH simulations.  
We  find that  the dark energy  modeling  has  an  unexpected  
significant  effect  on  galaxy formation, on the contrary of what is most 
commonly believed.

The experiment used a suite of SPH zoom-in cosmological simulations with masses of 
 $3.4\times10^{11}\Msun$, $6.6\times10^{11}\Msun$ and
$7.7\times10^{11}\Msun$, in four dynamical dark energy models plus the
reference  \LCDM model. 
The models all employed the \emph{same} baryonic physics prescription. All dynamical dark energy 
models lay in within the two sigma contours given by WMAP7. 
We examined the dark matter distribution, 
gas, star and total halo masses, star formation histories, baryonic matter 
distribution (rotation curves and surface brightness profiles), and the 
chemical enrichment of the galaxies.

Changing the dark energy evolution implies changing the expansion rate of
the Universe, which in turn affects the accretion history. We show how the same
galaxy evolved in different dark energy cosmologies does not present 
significant differences in dark matter only simulations, while in 
hydrodynamical simulations the galactic properties change greatly.

At $z=0$, the stellar mass inside $r_{vir}$ can vary by a factor of 2 depending 
on cosmology, while the dark matter mass only changes of a few
percent.  Thus baryons amplify differences
between dark  energy models, as the evolution of
the stellar mass - halo mass ratio shows: by simply changing the dark energy
parametrization stellar mass either decreases or increase of a
factor of two throughout the whole galaxy evolution.

The reason why baryons amplify the differences 
among the various dynamical dark energy models lays on the non linear 
response of the hydrodynamical processes. Once the cosmological model
introduces slightly different density perturbations, feedback processes enhances
those differences by producing slightly more (or less) stars.
More stars introduce more metals in the feedback cycle and more metals
decrease the cooling time, which in turn allows gas to cool faster and produce even 
more stars.
Through the highly non-linear response of baryons, dark energy models that would
have been indistinguishable from \LCDM on galactic scales show distinctive
features in SPH simulations.

The distinctive features of dynamical dark energy become clear when looking at
the star formation rates. We find that certain dark energy models are
able  to  both   delay  and  suppress  star   formation  until  recent
epochs. The delay  in star formation is then in  turn responsible for
the flattening  of rotations curves, where  we show a change of about
100 km/s  in the  two most massive galaxies  we considered.  Throughout the
analysis,  the least  massive galaxy is the least sensitive to  dark
energy  parametrization  changes,  in  agreement with  the  fact  that stellar
feedback is most effective around $10^{11}\Msun$ (\citealt{Dic14}).
The two most massive galaxies living in a 
slower expanding universe (waCDM2 model) have steeper surface brightness 
profiles due to their delayed star formation.

Finally  we compare the effect  of  dynamical dark  energy with  the
effect  of baryonic feedback.  We keep the cosmology fixed  (waCDM2) and
change the feedback  parametrization.  Provided  that the
Supernova feedback is kept constant, at late times the effect of dark
energy  is comparable  to the  effect of  early stellar feedback (see
\citealt{Sti13} for details on feedback modeling). Even the degree
at  which stellar  feedback is  able  to flatten  rotation curves,  is
comparable to the  effect of dark energy. We noted  on the other hand,
that in order  to obtain the behavior suggested  by abundance matching
considerations at  high redshifts,  early stellar  feedback had  to be
introduced since at high redshifts it has the most important effect 
compared to the dark energy modeling.

Having shown that the dark energy  modeling has an important effect on
galaxy formation  and evolution, we would  like to stress on  the fact
that, especially in  the era of high precision  cosmology, the details
on dark energy do matter and certainly need further investigations.

\section*{Acknowledgments} 
The analysis was performed using the pynbody package (\texttt{http://code.google.com/p/pynbody}),
which was written by Andrew Pontzen and Rok Ro\v{s}kar in addition to the authors.
Luciano Casarini acknowledges the Brazilian research Institutions FAPES and CNPq for financial support.\\
The numerical simulations used in this work were performed on the THEO
cluster   of  the   Max-Planck-Institut   f\"ur   Astronomie  at   the
Rechenzentrum in Garching. And A.M. would like to acknowledge the support from the 
Sonderforschungsbereich SFB 881 "the Milky Way System" of the German Research Foundation (DFG).
Greg Stinson received funding from the 
European Research Council under the European Union's 
Seventh Framework Programme (FP 7)  ERC Grant Agreement n. [321035].

\bibliographystyle{apj} \bibliography{biblio}

\bsp
\label{lastpage}
\end{document}